\documentclass[conference,comsoc]{IEEEtran}
\IEEEoverridecommandlockouts
\usepackage{cite}
\usepackage{amsmath,amssymb,amsfonts}
\usepackage{algorithmic}
\usepackage{graphicx}
\usepackage{textcomp}
\usepackage{xcolor}

\usepackage[T1]{fontenc}
\usepackage{graphicx}
\usepackage{amsmath}
\usepackage{bm}
\usepackage[cmintegrals]{newtxmath}

\def\BibTeX{{\rm B\kern-.05em{\sc i\kern-.025em b}\kern-.08em
    T\kern-.1667em\lower.7ex\hbox{E}\kern-.125emX}}
\begin{document}

\title{Opening the Black Box of Deep Neural Networks in Physical Layer Communication\\

\thanks{This work was supported in part by National Natural Science Foundation of China (NSFC) under Grant 61931020, 61372099 and 61601480.}
}

\author{Jun Liu,~Haitao Zhao,~Dongtang Ma,~Kai Mei~and~Jibo Wei\\
	\emph{College of Electronic Science and Technology}\\
	National University of Defense Technology,~Changsha~410073,~China\\
	E-mail:~\{liujun15,~haitaozhao,~dongtangma,~meikai11,~wjbhw\}@nudt.edu.cn}

\maketitle

\begin{abstract}
Deep Neural Network (DNN)-based physical layer techniques are attracting considerable interest due to their potential to enhance communication systems. However, most studies in the physical layer have tended to focus on the application of DNN models to wireless communication problems but not to theoretically understand how does a DNN work in a communication system. In this paper, we aim to quantitatively analyze why DNNs can achieve comparable performance in the physical layer comparing with traditional techniques and their cost in terms of computational complexity. We further investigate and also experimentally validate how information is flown in a DNN-based communication system under the information theoretic concepts.
\end{abstract}

\begin{IEEEkeywords}
Deep neural network (DNN), physical layer communication, information theory.
\end{IEEEkeywords}

\section{Introduction}
\label{Introduction}
%
%
%
%
\IEEEPARstart{D}{eep} neural networks (DNN) have recently drawn a lot of attention as a powerful tool in science and engineering problems such as protein structure prediction, image recognition, speech recognition and natural language processing that are virtually impossible to explicitly formulate. Although the mathematical theories of communication systems have been developed dramatically since Claude Elwood Shannon's monograph ``A mathematical theory of communication'' \cite{shannon1948mathematical} provides the foundation of digital communication, the wireless channel-related gap between theory and practice motivates researchers to implement DNNs in existing physical layer communication. In order to mitigate the gap, a natural thought is to let a DNN to jointly optimize a transmitter and a receiver for a given channel model without being limited to component-wise optimization. In \cite{o2017introduction}, a pure data-driven end-to-end communication system is proposed to jointly optimize transmitter and receiver components. Then, the authors consider the linear and nonlinear steps of processing the received signal as a radio transformer network (RTN) which can be integrated into the end-to-end training process. The ideas of end-to-end learning of communication system and RTN through DNN are extended to orthogonal frequency division multiplexing (OFDM) in \cite{felix2018ofdm}. Another natural idea is to recover channel state information (CSI) and estimate the channel as accurate as possible by implementing a DNN so that the effects of fading could be reduced. The authors of \cite{wen2018deep} propose an end-to-end DNN-based CSI compression feedback and recovery mechanism which is further extended with long short-term memory (LSTM) \cite{wang2018deep}. In \cite{li2019deep}, a residual learning based DNN designed for OFDM channel estimation is introduced. Furthermore, in order to mitigate the disturbances, in addition to Gaussian noise, such as channel fading and nonlinear distortion, \cite{liu2019online} proposes an online fully complex extreme learning machine-based symbol detection scheme.

Comparing with traditional physical layer communication systems, the above-mentioned DNN-based techniques show competitive performance. However, what has been missing is to understand the dynamics behind the DNN in physical layer communication. 

In this paper, we attempt to first give a mathematical explanation to reveal the mechanism of end-to-end DNN-based communication systems. Then, we try to unveil the role of the DNNs in the tasks of CSI recovery, channel estimation and symbol detection. We believe that we have developed a concise way to open as well as understand the ``black box'' of DNNs in physical layer communication. To summarize, our main contributions of this paper are twofold:
\begin{itemize}
	\item Instead of proposing a scheme combining a DNN with a typical communication system, we analyze the behaviours of a DNN-based communication system from the perspectives of the whole DNN (communication system), encoder (transmitter) and decoder (receiver). Our simulation results verify that the constellations produced by autoencoders are equivalent to the (locally) optimum constellations obtained by the gradient-search algorithm which minimize the asymptotic probability of error in Gaussian noise under an average power constraint.
	\item We consider the tasks of CSI recovery, channel estimation and symbol detection as a typical inference problem. The information flow in the DNNs of these tasks is estimated by using matrix-based functional of Renyi’s $\alpha$-entropy to approximate Shannon’s entropy.
\end{itemize}

The remainder of this paper is organized as follows. In Section \ref{System Model and Problem Formulation}, we give the system model and formulate the problem. Next, simulation results are presented in Section \ref{Simulation Results}. Finally, the conclusions are drawn in Section \ref{Conclusion}.

\textit{Notations:}~The notations adopted in the paper are as follows. We use boldface lowercase $\bf{x}$, capital letters $\bf{X}$ and calligraphic letters $\mathcal X$ to denote column vectors, matrices and sets respectively. In addition,~$  \odot  $ and~${\mathbb E}\left[  \cdot  \right]$ denote respectively the Hadamard product and the expectation operation.

\section{System Model and Problem Formulation}
\label{System Model and Problem Formulation}

\begin{figure}[t]
	\centering
	\includegraphics[width=3.6in]{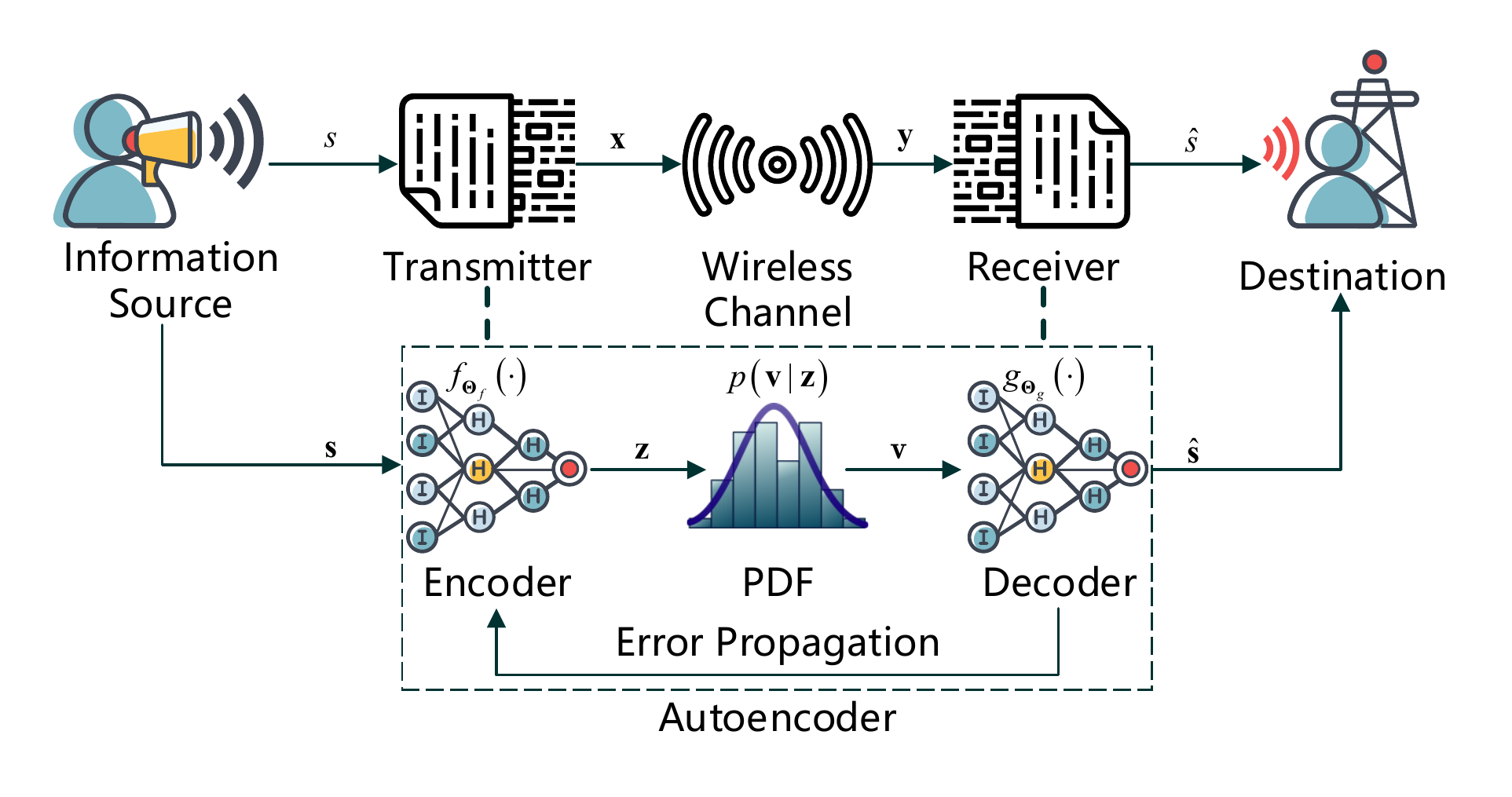}
	\caption{Schematic diagram of a general communication system and its corresponding autoencoder representation.}
	\label{CommunicationsSystems}
\end{figure}
In this section, we first describe the considered system model and then provide a detailed explanation of the problem formulation from three different perspectives. 

\subsection{System Model}
As shown in the upper part of Fig. \ref{CommunicationsSystems}, let's consider the process of message transmission from the perspectives of a typical communication system and an autoencoder, respectively. We assume that an information source generates a sequence of $k$-bit message symbols ${s} \in \left\{ {1,2, \cdots ,M} \right\}$ to be communicated to the destination, where $M=2^k$. Then the modulation modules inside the transmitter map each symbol $s$ to a signal ${\bf{x}} \in {\mathbb{R}^N}$, where $N$ denoted the dimension of the signal space. The signal alphabet is denoted by ${{\bf{x}}_1},{{\bf{x}}_2}, \cdots ,{{\bf{x}}_M}$. During channel transmission, $N$-dimensional signal $\bf{x}$ is corrupted to ${\bf{y}} \in {\mathbb{R}^N}$ with conditional probability density function (PDF) $p\left( {{\bf{y}}|{\bf{x}}} \right) = \prod _{n = 1}^Np\left( {{y_n}|{x_n}} \right)$. In this paper, we use $N/2$ bandpass channels, each with separately modulated inphase and quadrature components to transmit the $N$-dimensional signal \cite{sklar2014digital}. Finally, the received signal is mapped by the demodulation module inside the receiver to ${\hat s}$ which is an estimate of the transmitted symbol $s$. The procedures mentioned above have been exhaustively presented by Shannon.

From the point of view of filtering or signal inference, the idea of autoencoder-based communication system matches Norbert Wiener's perspective \cite{yu2017autoencoders}. As shown in the lower part of the Fig. \ref{CommunicationsSystems}, the autoencoder consists of an encoder and a decoder and each of them is a feedforward neural network (NN) with parameters (weights and biases) ${{{\bf{\Theta }}_f}}$ and ${{{\bf{\Theta }}_g}}$, respectively. Note that each symbol $s$ from information source usually needs to be encoded to a one-hot vector ${\bf{s}} \in {\mathbb{R}^M}$ and then is fed into the encoder. Under a given constraint (e.g., average signal power constraint), the PDF of a wireless channel and a loss function to minimize error symbol probability, the encoder and decoder are respectively able to learn to appropriately represent $\bf{s}$ as ${\bf{z}} = {f_{{{\bf{\Theta }}_f}}}\left( {\bf{s}} \right)$ and to map the corrupted signal $\bf{v}$ to an estimate of transmitted symbol ${\bf{\hat s}} = {g_{{{\bf{\Theta }}_g}}}\left( {\bf{v}} \right)$ where ${\bf{z}},{\bf{v}} \in {\mathbb{R}^N}$. Here, we use ${{\bf{z}}_1},{{\bf{z}}_2}, \cdots ,{{\bf{z}}_M}$ denoted the transmitted signals from the encoder in order to distinguish it from the transmitted signals from the transmitter.

\subsection{Understanding Autoencoders on Message Transmission}
From the prospective of the whole autoencoder (communication system), it aims to transmit information to destination with low error probability. The symbol error probability, i.e., the probability that the wireless channel has shifted a signal point into another signal's decision region, is 
\begin{equation}
	{P_e} = \frac{1}{M}\sum\limits_{m = 1}^M {\Pr \left( {{\bf{\hat s}} \ne {{\bf{s}}_m}|{{\bf{s}}_m}~{\rm{transmitted}}} \right)}.
	\label{Pe}
\end{equation}
The autoencoder can use the cross-entropy loss function
\begin{equation}
	\begin{aligned}
		{\mathcal L_{\log }}\left( {{\bf{\hat s}},{\bf{s}};{{\bf{\Theta }}_f},{{\bf{\Theta }}_g}} \right) &=  - \frac{1}{B}\sum\limits_{b = 1}^B {\sum\limits_{i = 1}^M {{{\bf{s}}^{\left( b \right)}}\left[ i \right]\log \left( {{{{\bf{\hat s}}}^{\left( b \right)}}\left[ i \right]} \right)} } \\ 
		&  =  - \frac{1}{B}\sum\limits_{b = 1}^B {\log \left( {{{{\bf{\hat s}}}^{\left( b \right)}}\left[ s \right]} \right)}  
	\end{aligned}
	\label{LossFunction}
\end{equation}
to represent the price paid for inaccuracy of prediction where ${{{\bf{s}}^{\left( b \right)}}\left[ i \right]}$ denotes the $i$-th element of the $b$-th symbol in a training set with $B$ symbols. In order to train the autoencoder to minimize the symbol error probability, the optimal parameters could be found by optimizing the loss function
\begin{equation}
	\begin{aligned}
		&\left( {{\bf{\Theta }}_f^ * ,{\bf{\Theta }}_g^ * } \right) = \mathop {\arg \min }\limits_{\left( {{{\bf{\Theta }}_f},{{\bf{\Theta }}_g}} \right)} \left[ {{\mathcal L_{\log }}\left( {{\bf{\hat s}},{\bf{s}};{{\bf{\Theta }}_f},{{\bf{\Theta }}_g}} \right)} \right]\\
		&{\rm{subject~to~}}{\mathbb E}\left[ {\left\| {\bf{z}} \right\|_2^2} \right] \le {P_{{\rm{av}}}}
	\end{aligned}
	\label{Optimization_AE}
\end{equation}
where $P_{{\rm{av}}}$ denotes the average power. In this paper, we set $P_{{\rm{av}}}={1/M}.$ Now, we must be very curious about how does the mapping ${\bf{z}} = {f_{{{\bf{\Theta }}_f}}}\left( {\bf{s}} \right)$  look like after the training was done.

\subsection{Encoder: Finding a Good Representation}
Let's pay attention to the encoder (transmitter). In the domain of communication, an encoder needs to learn a robust representation ${\bf{z}} = {f_{{{\bf{\Theta }}_f}}}\left( {\bf{s}} \right)$ to transmit $\bf{s}$ against channel disturbances, such as thermal noise, channel fading, nonlinear distortion, phase jitter, etc. This is equivalent to find a coded (or uncoded) modulation scheme with the signal set of size $M$ to map a symbol $\bf{s}$ to a point $\bf{z}$ for a given transmitted power, which maximizes the minimum distance between any two constellation points. Usually the problem of finding good signal constellations for a Gaussian channel\footnote{The problem of constellation optimization is usually considered under the condition of the Gaussian channel. Although the problem under the condition of Rayleigh fading channel has been studied in \cite{boutros1996good}, its prerequisite is that the side information is perfect known.} is associated with the search for lattices with high packing density which is an old and well-studied problem in the mathematical literature \cite{jorge2015algebraic}.

We use the algorithm proposed in \cite{foschini1974optimization} to obtain the optimum constellations. Consider a zero-mean stationary additive white Gaussian noise (AWGN) channel with one-sided spectral density $2N_0$. For large signal-to-noise ratio (SNR), the asymptotic approximation of the (\ref{Pe}) can be written as 
\begin{equation}
	{P_e} \sim \exp \left( { - \frac{1}{{8{N_0}}}\mathop {\min }\limits_{i \ne j} \left\| {{{\bf{z}}_i} - {{\bf{z}}_j}} \right\|_2^2} \right).
\end{equation}
To minimize $P_e$, the problem can be formulated  as
\begin{equation}
	\begin{aligned}
		&\left\{ {{\bf{z}}_m^ * } \right\}_{m = 1}^M = \mathop {\arg \min }\limits_{\left\{ {{{\bf{z}}_m}} \right\}_{m = 1}^M} \left( {{P_e}} \right)\\
		&{\rm{subject~to~}}{\mathbb E}\left[ {\left\| {\bf{z}} \right\|_2^2} \right] \le {P_{{\rm{av}}}}
	\end{aligned}
	\label{Optimization_Asymptotic}
\end{equation}
where $\left\{ {{\bf{z}}_m^ * } \right\}_{m = 1}^M$ denotes the optimal signal set. This optimization problem can be solved by using a constrained gradient-search algorithm. We denote $\left\{ {{\bf{z}}_m } \right\}_{m = 1}^M$ as an $M \times N$ matrix
\begin{equation}
	{\bf{Z}} = {\left[ {{{\bf{z}}_1},{{\bf{z}}_2}, \cdots ,{{\bf{z}}_M}} \right]^T}.
\end{equation}
Then, the $s$-th step of the constrained gradient-search algorithm can be described by
\begin{subequations}
	\begin{align}
		&{\bf{Z}}_{s + 1}^{\rm{'}} = {{\bf{Z}}_s} - {\eta_s}\nabla {P_e}\left( {{{\bf{Z}}_s}} \right) \\
		&{{\bf{Z}}_{s + 1}} = \frac{{{\bf{Z}}_{s + 1}^{\rm{'}}}}{{\sum\limits_i {\sum\limits_j {{{\left( {{\bf{Z}}_{s + 1}^{\rm{'}}\left[ {i,j} \right]} \right)}^2}} } }}
	\end{align}
	\label{MthStep}
\end{subequations}
where $\eta_s$ denotes step size and $\nabla {P_e}\left( {{{\bf{Z}}_s}} \right) \in {\mathbb{R}^{M \times N}}$ denotes the gradient of $P_e$ respect to the current constellation points.
It can be written as
\begin{equation}
	\nabla {P_e}\left( {{{\bf{Z}}_s}} \right) = {\left[ {{{\bf{g}}_1},{{\bf{g}}_2}, \cdots ,{{\bf{g}}_M}} \right]^T}
\end{equation}
where
\begin{equation}
	\resizebox{1 \hsize}{!}{$
		{{\bf{g}}_m} \sim  - \sum\limits_{i \ne m} {\exp \left( { - \frac{{\left\| {{{\bf{z}}_m} - {{\bf{z}}_i}} \right\|_2^2}}{{8{N_0}}}} \right)\left( {\frac{1}{{\left\| {{{\bf{z}}_m} - {{\bf{z}}_i}} \right\|_2^2}} + \frac{1}{{4{N_0}}}} \right){{\bf{1}}_{{{\bf{z}}_m} - {{\bf{z}}_i}}}} 
	$}.
\end{equation}
The vector ${{{\bf{1}}_{{{\bf{z}}_m} - {{\bf{z}}_i}}}}$ denotes $N$-dimensional unit vector in the direction of ${{{\bf{z}}_m} - {{\bf{z}}_i}}$.

Comparing (\ref{Optimization_AE}) to (\ref{Optimization_Asymptotic}), the mechanism of the encoder in an autoencoder-based communication system has been unveiled. The mapping function of encoder can be represented as
\begin{equation}
	\left\{ {{f_{{\bf{\Phi }}_f^ * }}\left( {{{\bf{s}}_m}} \right)} \right\}_{m = 1}^M \to \left\{ {{\bf{z}}_m^ * } \right\}_{m = 1}^M
\end{equation}
when the PDF used for generating training samples is multivariate zero-mean normal distribution ${\bf{\hat z}} - {\bf{z}} \sim {{\cal N}_N}({\bf{\vec 0}},{\mkern 1mu} {\bm{\Sigma }})$ where ${{\bf{\vec 0}}}$ denotes $N$-dimensional zero vector and ${\bm{\Sigma }} = \left( {2{N_0}/N} \right){\bf{I}}$ is an $N \times N$ diagonal matrix.
\subsection{Decoder: Inference}
\begin{figure}[t]
	\centering
	\includegraphics[width=3.45in]{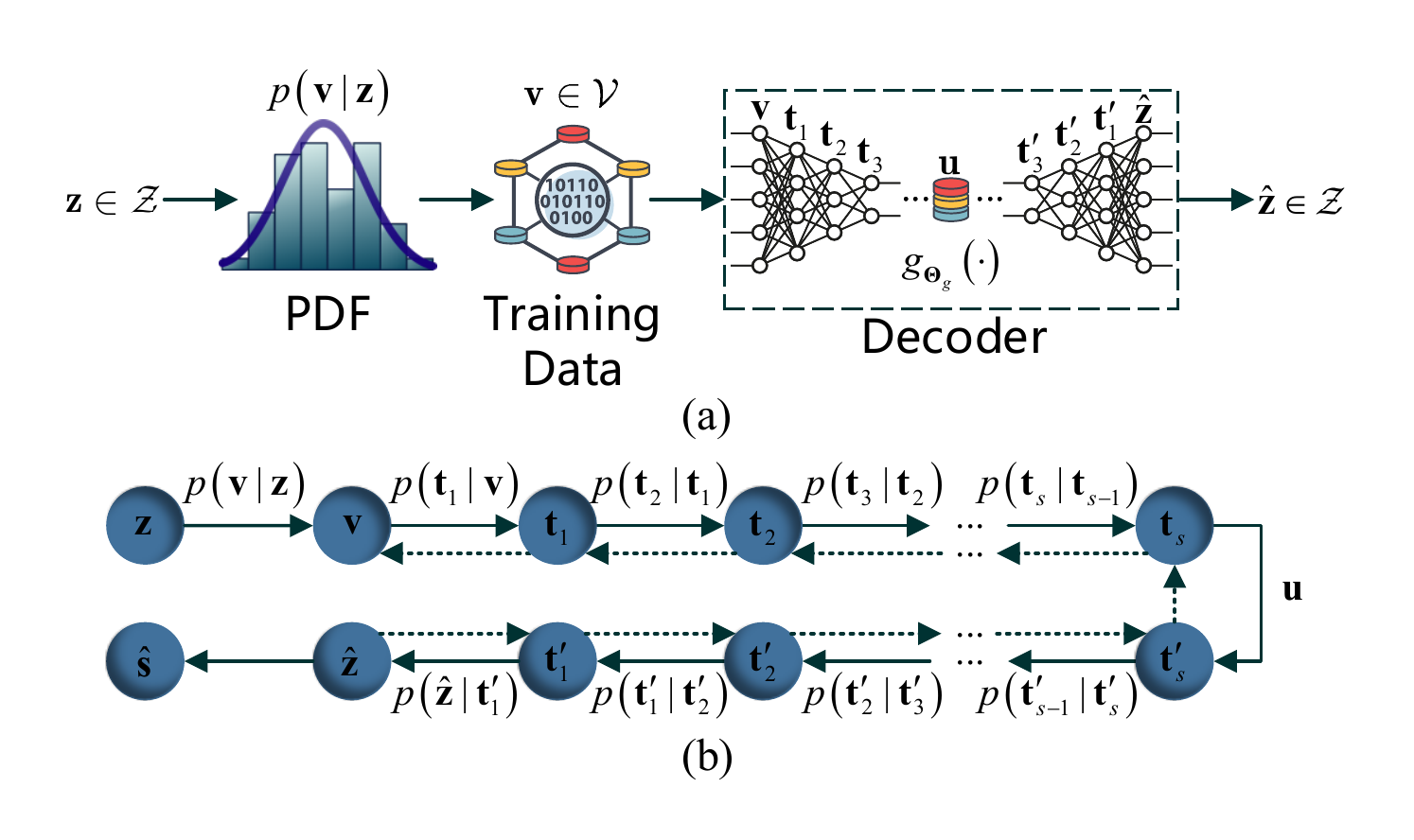}
	\caption{(a) An inference model with a DNN decoder of size $\left( {2S - 1} \right)$ hidden layers for learning. (b) The graph representation of the decoder with $\left( {S - 1} \right)$ hidden layers in both sub encoder and decoder. The solid arrow denotes the direction of input feedforward propagation and the dashed arrow denotes the direction of information flow in the error back-propagation phase.}
	\label{Decoder}
\end{figure}
Finally, it is the time to zoom in the lower right corner of the Fig. \ref{CommunicationsSystems} to investigate what happens inside the decoder (receiver). As Fig. \ref{Decoder}(a) shown, for the tasks of DNN-based CSI recovery, channel estimate and symbol detection, the problem can be formulated as an inference model. For the sake of convenience, we can denote the target output of the decoder as $\bf{z}$ instead of $\bf{s}$ because we can assume  ${\bf{z}} = {f_{{{\bf{\Theta }}_f}}}\left( {\bf{s}} \right)$ is bijection. If the decoder is symmetric, the decoder also can be seen as a sub autoencoder which consists of  a sub encoder and decoder. Its bottleneck (or middlemost) layer codes is denoted as $\bf{u}$. Here we use $\bf{z}$ to denote CSI or transmitted symbol which we desire to predict. The decoder infers a prediction ${\bf{\hat z}} = {g_{{{\bf{\Theta }}_g}}}\left( {\bf{v}} \right)$ according to its corresponding measurable variable $\bf{v}$.

If the joint distribution $p\left( {{\bf{v}},{\bf{z}}} \right)$ is known, the expected (population) risk ${{\cal C}_{p\left( {{\bf{v}},{\bf{z}}} \right)}}\left( {{g_{{{\bf{\Theta }}_g}}},{{\cal L}_{\log }}} \right)$ can be written as
\begin{equation}
	\begin{aligned}
		{\mathbb{E}}\left[ {{{\cal L}_{\log }}\left( {{\bf{\hat z}},{\bf{z}};{{\bf{\Theta }}_g}} \right)} \right] &= \sum\limits_{{\bf{v}} \in \mathcal V,{\bf{z}} \in \mathcal Z} {p\left( {{\bf{v}},{\bf{z}}} \right)\log \left( {\frac{1}{{Q\left( {{\bf{z}}|{\bf{v}}} \right)}}} \right)}\\
		&=\sum\limits_{{\bf{v}} \in \mathcal V,{\bf{z}} \in \mathcal Z} {p\left( {{\bf{v}},{\bf{z}}} \right)\log \left( {\frac{1}{{p\left( {{\bf{z}}|{\bf{v}}} \right)}}} \right)}{\rm{ + }}\\
		&~~~\sum\limits_{{\bf{v}} \in \mathcal V,{\bf{z}} \in \mathcal Z} {p\left( {{\bf{v}},{\bf{z}}} \right)\log \left( {\frac{{p\left( {{\bf{z}}|{\bf{v}}} \right)}}{{Q\left( {{\bf{z}}|{\bf{v}}} \right)}}} \right)}\\
		&=H\left( {{\bf{z}}|{\bf{v}}} \right) + {D_{{\rm{KL}}}}\left( {p\left( {{\bf{z}}|{\bf{v}}} \right)||Q\left( {{\bf{z}}|{\bf{v}}} \right)} \right)\\
		& \ge H\left( {{\bf{z}}|{\bf{v}}} \right) 
	\end{aligned}
	\label{PopulationRisk}
\end{equation}
where $Q\left( { \cdot |{\bf{v}}} \right){\rm{ = }}{g_{{{\bf{\Theta }}_g}}}\left( {\bf{v}} \right) \in p\left( {\mathcal{Z}} \right)$ and ${D_{{\rm{KL}}}}\left( {p\left( {{\bf{z}}|{\bf{v}}} \right)||Q\left( {{\bf{z}}|{\bf{v}}} \right)} \right)$ denotes Kullback-Leibler divergence between ${p\left( {{\bf{z}}|{\bf{v}}} \right)}$ and ${Q\left( {{\bf{z}}|{\bf{v}}} \right)}$ \cite{zaidi2020information}\footnote{If $\bf{z}$ and $\bf{v}$ are continuous random variables, the sum becomes an integral when their PDFs exist.}. If and only if the decoder is given by the conditional posterior ${g_{{{\bf{\Theta }}_g}}}\left( {\bf{v}} \right){\rm{ = }}p\left( {{\bf{z}}|{\bf{v}}} \right)$, the expected (population) risk reaches the minimum $\mathop {\min }\limits_{{g_{{{\bf{\Theta }}_g}}}} {{\cal C}_{p\left( {{\bf{v}},{\bf{z}}} \right)}}\left( {{g_{{{\bf{\Theta }}_g}}},{{\cal L}_{\log }}} \right) = H\left( {{\bf{z}}|{\bf{v}}} \right)$.

In physical layer communication, instead of perfectly knowing the channel-related joint distribution $p\left( {{\bf{v}},{\bf{z}}} \right)$, we only have a set of $B$ \textit{i.i.d.} samples ${{\cal D}_B}: = \left\{ {\left( {{{\bf{v}}^{\left( b \right)}},{{\bf{z}}^{\left( b \right)}}} \right)} \right\}_{b = 1}^B$ from $p\left( {{\bf{v}},{\bf{z}}} \right)$. In this case, the empirical risk is defined as
\begin{equation}
	{\hat {\cal C}_{p\left( {{\bf{v}},{\bf{z}}} \right)}}\left( {{g_{{{\bf{\Theta }}_g}}},{\cal L},{{\cal D}_B}} \right) = \frac{1}{B}\sum\limits_{b = 1}^B {{\cal L}\left[ {{{\bf{z}}_b},{g_{{{\bf{\Theta }}_g}}}\left( {{{\bf{v}}_b}} \right)} \right]}.
\end{equation}
Practically, the ${\mathcal D}_B$ from $p\left( {{\bf{v}},{\bf{z}}} \right)$ usually is a finite set. This leads the difference between the empirical and expected (population) risks which can be defined as
\begin{equation}
	\begin{aligned}
		{\rm{ge}}{{\rm{n}}_{p\left( {{\bf{v}},{\bf{z}}} \right)}}\left( {{g_{{{\bf{\Theta }}_g}}},{\cal L},{{\cal D}_B}} \right)=&{{\cal C}_{p\left( {{\bf{v}},{\bf{z}}} \right)}}\left( {{g_{{{\bf{\Theta }}_g}}},{{\cal L}_{\log }}} \right)-\\
		&{\hat {\cal C}_{p\left( {{\bf{v}},{\bf{z}}} \right)}}\left( {{g_{{{\bf{\Theta }}_g}}},{\cal L},{{\cal D}_B}} \right).
	\end{aligned}
\end{equation}

We now can preliminarily conclude that the DNN-based receiver is an estimator with minimum empirical risk for a given set ${\mathcal D}_B$, whereas its performance is inferior to the optimal with minimum expected (population) risk under a given joint distribution $p\left( {{\bf{v}},{\bf{z}}} \right)$.

Furthermore, it is necessary to quantitatively understand how information flows inside the decoder. Fig. \ref{Decoder}(b) shows the graph representation of the decoder where ${{{\bf{t}}_i}}$ and ${{{\bf{t'}}}_i}\left( {1 \le i \le S} \right)$ denote $i$-th hidden layer representations starting from the input layer and the output layer, respectively. Here, we use the method proposed in \cite{yu2019understanding} to illustrate layer-wise mutual information by three kinds of information planes (IPs) where the Shannon’s entropy is estimated by matrix-based functional of Renyi’s $\alpha$-entropy \cite{giraldo2014measures}. Its details are given in Appendix.

\section{Simulation Results}
\label{Simulation Results}
\begin{figure}[t]
	\centering
	\includegraphics[width=3.45in]{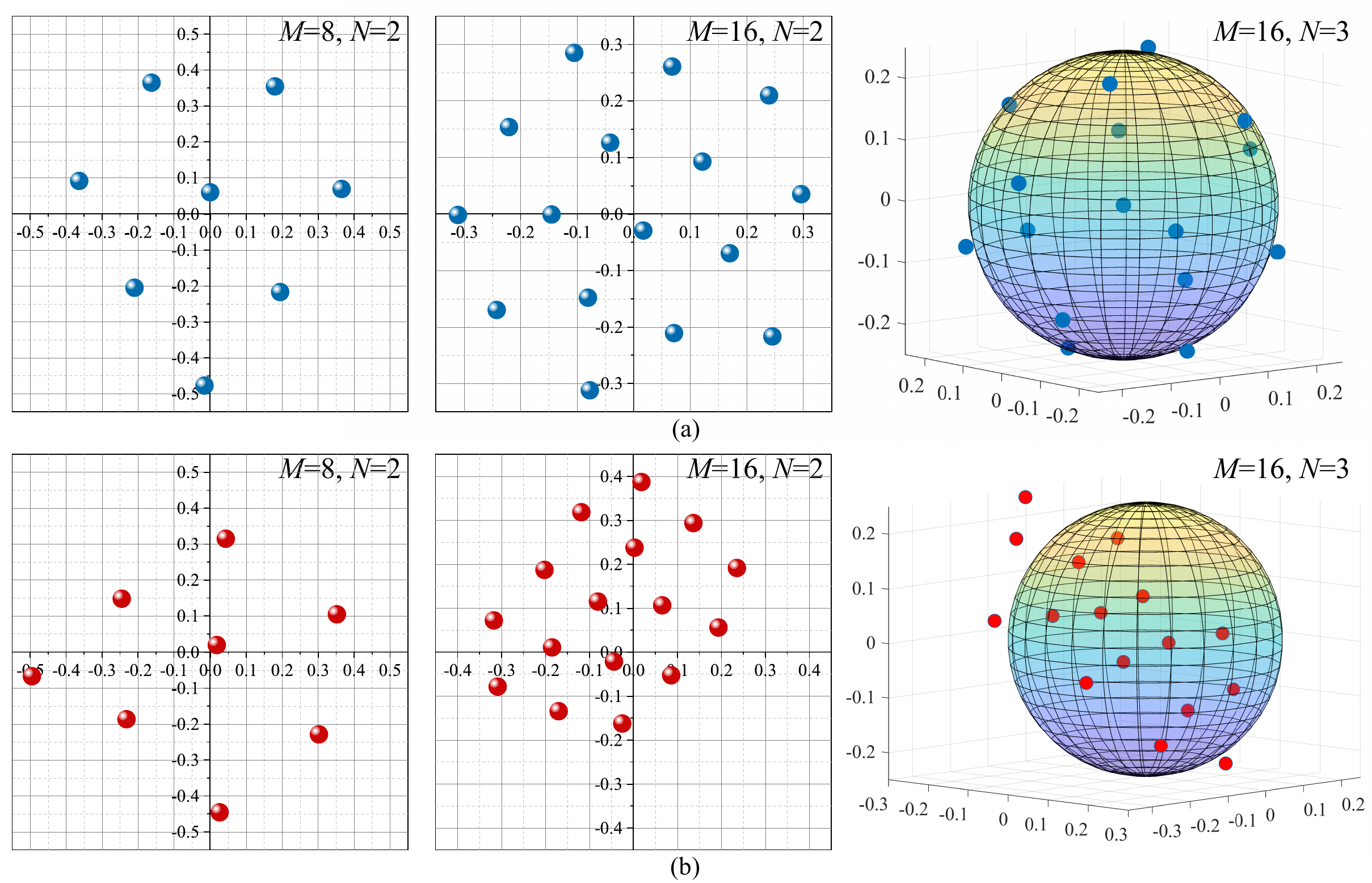}
	\caption{Comparisons of (a) optimum constellations obtained by gradient-search technique and (b) constellations produced by autoencoders.}
	\label{Constellations}
\end{figure}
In this section, we provide simulation results to illustrate the behavior of DNN in physical layer communication.
\subsection{Constellation Comparison}
Fig. \ref{Constellations}(a) shows the optimum constellations obtained by gradient-search technique proposed in \cite{foschini1974optimization}. When $N=2$ and $3$, the algorithm was run allowing for 1000 and 3000 steps, respectively. The step size $\eta  = 2 \times {10^{ - 4}}$. Fig. \ref{Constellations}(b) shows the constellations produced by autoencoders which have the same network structures and hyperparameters with the autoencoders mentioned in \cite{o2017introduction}. The autoencoders were trained with ${10^6}$ epochs, each of which contains $M$ different symbols. 

When $N=2$, the two-dimensional constellations produced by autoencoders have a similar pattern to the optimum constellations which form a lattice of (almost) hexagonal. Specifically, in the case of $\left( {M = 8,\;N = 2} \right)$, one of the constellations found by the autoencoder can be obtained by rotating the optimum constellation found by gradient-search technique. In the case of $\left( {M = 16,\;N = 2} \right)$, the constellation found by the autoencoder is different from the optimum constellation but it still forms a lattice of (almost)  equilateral triangles. In the case of $\left( {M = 16,~N = 3} \right)$, one signal point of the optimum constellation is almost at the origin while the other 15 signal points are almost at the surface of a sphere with radius $P_{{\rm{av}}}$ and center 0. This pattern is similar to the surface of a truncated icosahedron which is composed of pentagonal and hexagonal faces. However, the three-dimensional constellation produced by an autoencoder is a local optima which is form by 16 signal points almost in a plane.  

From the perspective of computational complexity, the cost to train an autoencoder is significantly higher than the cost of traditional algorithm. Specifically, an autoencoder which has 4 dense layers respectively with $M$, $N$, $M$ and $M$ neurons needs to train $\left( {2M + 1} \right)\left( {M + N} \right) + 2M$ parameters for ${10^6}$ epochs whereas the gradient-search algorithm only needs $2M$ trainable parameters for ${10^3}$ steps.
\subsection{Information Flow}
We consider a common channel estimation problem for an OFDM system with $N$ subcarriers. Let ${\bf{z}} \buildrel \Delta \over = {\left[ {H\left[ 0 \right],H\left[ 1 \right], \cdots ,H\left[ {N - 1} \right]} \right]^T}$ which denotes frequency impulse response (FIR) vector of a channel. For the sake of convenience, we denote the measurable variable as ${\bf{v}} \buildrel \Delta \over = {{\bf{\hat z}}_{{\rm{LS}}}}$ where ${{\bf{\hat z}}_{{\rm{LS}}}}$ represents the least-square (LS) estimation of $\bf{z}$. Usually, it can be obtained by using training symbol-based channel estimation. In this paper, we use linear interpolation and the number of pilots ${N_p} = N/4=16$.

\begin{figure}[t]
	\centering
	\includegraphics[width=3.45in]{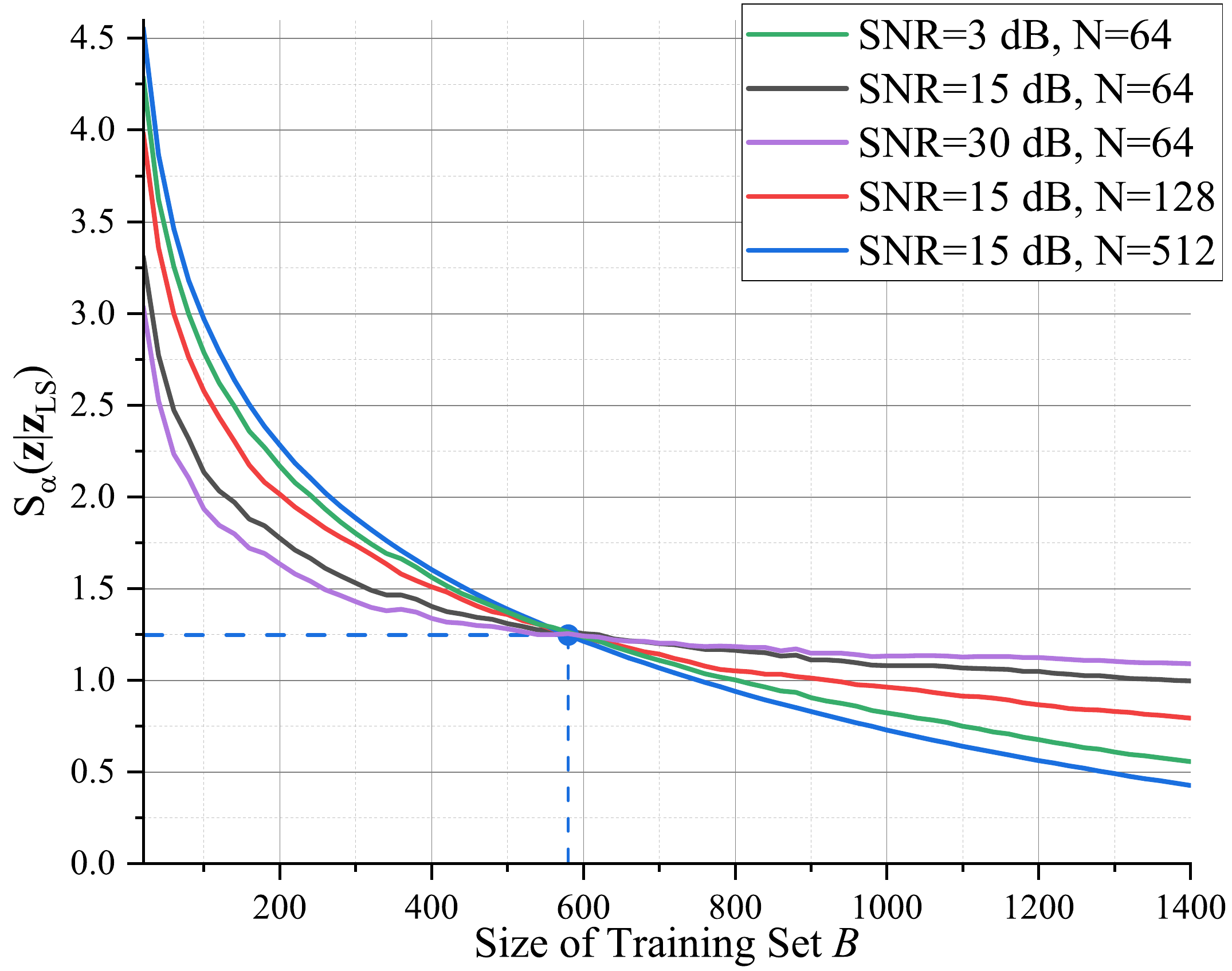}
	\caption{The entropy ${S_\alpha }\left( {{\bf{z}}|{{{\bf{\hat z}}}_{{\rm{LS}}}}} \right)$  with respect to different values of SNR and $N$.}
	\label{PopulationRiskFig}
\end{figure}

According to (\ref{PopulationRisk}), the minimum logarithmic expected (population) risk for this inference problem is $H\left( {{\bf{z}}|{{{\bf{\hat z}}}_{{\rm{LS}}}}} \right)$ which can be estimated by Renyi's $\alpha$-entropy ${S_\alpha }\left( {{\bf{z}}|{{{\bf{\hat z}}}_{{\rm{LS}}}}} \right){\rm{ = }}{S_\alpha }\left( {{\bf{z}},{{{\bf{\hat z}}}_{{\rm{LS}}}}} \right) - {S_\alpha }\left( {{{{\bf{\hat z}}}_{{\rm{LS}}}}} \right)$ with $\alpha=1.01$. Fig. \ref{PopulationRiskFig} illustrates the entropy ${S_\alpha }\left( {{\bf{z}}|{{{\bf{\hat z}}}_{{\rm{LS}}}}} \right)$ with respect to different values of SNR and $N$. As can be seen, ${S_\alpha }\left( {{\bf{z}}|{{{\bf{\hat z}}}_{{\rm{LS}}}}} \right)$ monotonically decreases as the size of training set increases. When $B \to \infty$, ${S_\alpha }\left( {{\bf{z}}|{{{\bf{\hat z}}}_{{\rm{LS}}}}} \right)$ decreases slowly. It is because the joint distribution $p\left( {{\bf{z}},{{{\bf{\hat z}}}_{{\rm{LS}}}}} \right)$ can be perfectly learned and therefore the empirical risk is approaching to the expected risk. Interestingly, when $B>580$, the lower the SNR or the larger input dimension $N$ is, the smaller $B$ is needed to obtain the same value of ${S_\alpha }\left( {{\bf{z}}|{{{\bf{\hat z}}}_{{\rm{LS}}}}} \right)$.

\begin{figure}[t]
	\centering
	\includegraphics[width=3.45in]{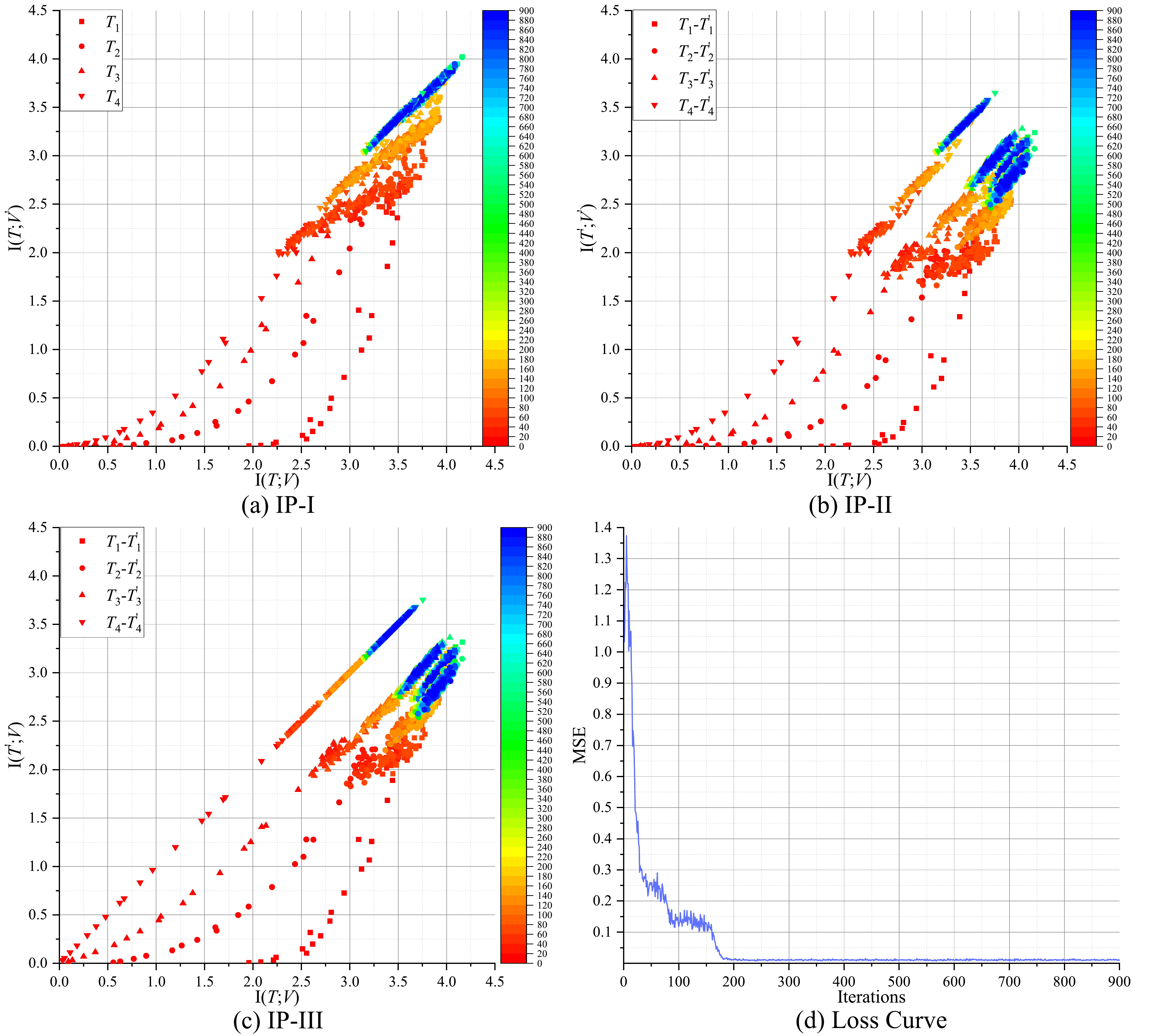}
	\caption{The three IPs and loss curve in a DNN-based channel estimator.}
	\label{InformationPlane}
\end{figure}

Fig. \ref{InformationPlane}(a), (b) and (c) illustrate the behavior of the IP-I, IP-II and IP-III  in a DNN-based OFDM channel estimator with topology ``$128 - 64 - 32 - 16 - 8 - 16 - 32 - 64 - 128$'' where linear activation function is considered and the training sample is constructed by concatenating the real and imaginary parts of the complex channel vectors. Batch size is 100 and learning rate $\eta=0.001$. We use $V$ and ${V'}$ to denote the input and output of the decoder, respectively. The number of iterations is illustrated through a color bar. From IP-I, it can be seen that the final value of mutual information ${\rm{I}}\left( {T;V'} \right)$ in each layer tends to be equal to the final value of ${\rm{I}}\left( {T;V} \right)$, which means that the information from $V$ has been learnt and transferred to $V'$ by each layer. In IP-II, ${\rm{I}}\left( {T';V'} \right) < {\rm{I}}\left( {T;V} \right)$ in each layer, which implies that all the layers are not overfitting. The tendency of ${\rm{I}}\left( {T;V} \right)$ to approach the value of ${\rm{I}}\left( {T';V} \right)$ can be observed from IP-III. Finally, from all the IPs, it is easy to notice that the mutual information does not change significantly when the number of iterations is larger than 200. Meanwhile, according to Fig. \ref{InformationPlane}(d), the MSE reaches a very low value and also does not change sharply. It means that 200 iterations are enough for the task of 64-subcarrier channel estimation using a DNN with the above-mentioned topology.

\section{Conclusion}
\label{Conclusion}
In this paper, we propose a framework to understand the manner of the DNNs in physical communication. We find that a DNN-based transmitter essentially tries to produce a good representation of the information source. Then, we quantitatively analyze the information flow in a DNN-based communication system. We believe that this framework has the potential for the design of DNN-based physical communication.

\appendices
\section{Matrix-based Functional of Renyi’s $\alpha$-Entropy}
For a random variable $X$ in a finite set $\mathcal X$, its Renyi’s entropy of order $\alpha$ is defined as
\begin{equation}
	{H_\alpha }\left( X \right) = \frac{1}{{1 - \alpha }}\log \int_{\cal X} {{f^\alpha }\left( x \right)dx} 
\end{equation}
where $f\left( x \right)$ is the PDF of the random variable $X$. Let $\left\{ {{x^{\left( b \right)}}} \right\}_{b = 1}^B$ be an \textit{i.i.d.} sample of $B$ realizations from the random variable $X$ with PDF $f\left( x \right)$. The Gram matrix ${\bf{K}}$ can be defined as ${\bf{K}}\left[ {i,j} \right] = \kappa \left( {{x_i},{x_j}} \right)$ where $\kappa :{\mathcal X} \times {\mathcal X} \mapsto {\mathbb R}$ is a real valued positive definite and infinitely divisible kernel. Then, a matrix-based analogue to Renyi’s $\alpha$-entropy for a normalized positive definite matrix $\bf{A}$ of size $B \times B$ with trace 1 can be given by the functional 
\begin{equation}
	{S_\alpha }\left( {\bf{A}} \right) = \frac{1}{{1 - \alpha }}{\log _2}\left[ {\sum\limits_{b = 1}^B {{\lambda _b}{{\left( {\bf{A}} \right)}^\alpha }} } \right]
\end{equation}
where ${{\lambda _b}\left( {\bf{A}} \right)}$ denotes the $b$-th eigenvalue of $\bf{A}$, a normalized version of $\bf{K}$:
\begin{equation}
	{\bf{A}}\left[ {i,j} \right] = \frac{1}{B}\frac{{{\bf{K}}\left[ {i,j} \right]}}{{\sqrt {{\bf{K}}\left[ {i,i} \right]{\bf{K}}\left[ {j,j} \right]} }}.
\end{equation}
Now, the joint-entropy can be defined as
\begin{equation}
	{S_\alpha }\left( {{\bf{A}},{\bf{B}}} \right) = {S_\alpha }\left[ {\frac{{{\bf{A}} \odot {\bf{B}}}}{{{\rm{tr}}\left( {{\bf{A}} \odot {\bf{B}}} \right)}}} \right].
\end{equation}
Finally, the matrix notion of Renyi’s mutual information can be defined as 
\begin{equation}
	{I_\alpha }\left( {{\bf{A}};{\bf{B}}} \right) = {S_\alpha }\left( {\bf{A}} \right) + {S_\alpha }\left( {\bf{B}} \right) - {S_\alpha }\left( {{\bf{A}},{\bf{B}}} \right).
\end{equation}



\ifCLASSOPTIONcaptionsoff
\newpage
\fi



%

%
%

\bibliographystyle{IEEEtran}  
\bibliography{IEEEabrv,Bibliography/MyCollection}

%








\end{document}